\documentclass{raa}            
\usepackage{graphicx,times}

\begin{document}

\title{The radio environment of the 21 Centimeter Array: 
RFI detection and mitigation}

   \volnopage{Vol.16 (2016) No.02 000--000}      
   \setcounter{page}{1}          

\author{Yan Huang\inst{1}
       \and Xiang-Ping Wu\inst{1}
       \and Qian Zheng\inst{2}         
       \and Jun-Hua Gu\inst{1} 
       \and Haiguang Xu\inst{3} 
}
\institute{National Astronomical Observatories, 
Chinese Academy of Sciences, Beijing 100012, China; {\it huangyan@bao.ac.cn}\\
\and
School of Chemical and Physical Sciences, 
Victoria University of Wellington, Wellington, New Zealand
\and
Department of Physics and Astronomy, Shanghai Jiao Tong University, 
Shanghai 200240, China
}

  \date{Received~~2015 11 23; accepted~~2015~~12 5}

\abstract{
Detection and mitigation of radio frequency interference (RFI) is the first 
and also the key step for data processing in radio observations, 
especially for ongoing low frequency radio experiments towards the 
detection of the cosmic dawn and epoch of reionization (EoR). 
In this paper we demonstrate the technique and efficiency of RFI 
identification and mitigation for the 21 Centimeter Array (21CMA), 
a radio interferometer dedicated to the statistical measurement of EoR. 
For terrestrial, man-made RFI, we concentrate mainly on a statistical 
approach by identifying and then excising 
non-Gaussian signatures, in the sense that the extremely weak cosmic 
signal is actually buried under thermal and therefore Gaussian noise. 
We also introduce the so-called visibility 
correlation coefficient instead of conventional visibility, which allows a 
further suppression of rapidly time-varying RFI. Finally, we briefly discuss 
removals of the sky RFI, the leakage of sidelobes from off-field strong 
radio sources with time-invariant power and a featureless spectrum. 
It turns out that state-of-the-art technique 
should allow us to detect and mitigate RFI to a satisfactory level 
in present low frequency interferometer observations such as those 
acquired with the 21CMA, and the accuracy and efficiency can be greatly 
improved with the employment of low-cost, high-speed computing facilities 
for data acquisition and processing. }


\authorrunning{Y. Huang, X.-P. Wu, Q. Zheng, J.-H. Gu \& H. Xu} 

\titlerunning{RFI in the 21CMA}

\maketitle

\section{Introduction}
\label{sect:intro}

Rapid progress in low frequency radio astronomy has been made over the 
last decade. This is primarily driven by the ambitious goal 
related to exploration of the cosmic dawn and epoch of reionization (EoR). 
Indeed, the probe of the dark ages, cosmic dawn and EoR will constitute 
the last frontier for observational 
cosmology in the era of precision cosmology. One of the most desirable 
goals is to unveil the history of how and when the universe was illuminated by 
the first stars/black holes and underwent a transition from its dark to 
bright phase. Thanks to the 21 cm radiation of neutral hydrogen, 
the most abundant type of baryonic matter in the early universe, 
we should be able to receive information about that important epoch of 
cosmic evolution, though the signal is extremely weak ($\sim1-10$ mK). 
Yet, the wavelength of the HI 21 cm radiation has been stretched by 
a factor of (1+$z$) due to the expansion of the universe. 
Therefore, if the cosmic dawn and EoR occurred 
at redshifts between 6 and 27 in terms of current observational 
constraints (for a recent summary see Koopmans et al. ~\cite{Koopmans15}), 
the corresponding HI hyperfine radiation has already been 
shifted to 1.5-6 m in wavelength or 50-200 MHz in frequency. Such 
a frequency band for us is common because FM radio 
is broadcast at 88-108 MHz. On one hand, radio experiments that 
probe the first lights from cosmic dawn and EoR have to work with 
low frequency radio astronomy if the HI 21 cm radiation is the unique 
message carrier. On the other hand, man-made radio 
frequency interference (RFI) will be a disaster in this frequency 
range. A remote site is the minimum requirement for low frequency 
radio observations that can be used to investigate the cosmic dawn and EoR.   

Technically, current low frequency radio experiments  have benefited from 
the rapid development of information technology and computer science: 
For example, conventional analog signal communication in radio astronomy has  
been replaced by high precision digital processing. A large amount of 
correlation computation and data communication between antennas in 
radio interferometers can be carried out by advanced, intelligent 
computers and networks. Massive amounts of data can be easily stored, 
transferred, shared and analyzed. All of these novel technologies have 
revolutionized traditional studies in radio astronomy including methods 
used for global VLBI. Instead of giant dish antennas, a large number of small 
antenna units can be integrated and combined to form a large 
collecting area, which can significantly increase the telescope's sensitivity 
while maintaining a large field of view. Equipped further with 
smart receivers, precise, high speed analog-to-digital converters, 
digital multi-beam forming, powerful computers and state-of-the-art 
data processing techniques, design and construction of 
radio telescopes are now moving from traditional 
mechanization and automation to modern digitization and intelligentization.  
Several new digital radio telescopes that incorporate advanced software 
have been built and served as the pathfinders or precursors for the ultimate 
radio telescope, the Square Kilometre Array 
(SKA)\footnote{http://www.skatelescope.org}, including 
GMRT\footnote{http://www.ncra.tifr.res.in}, 
LOFAR\footnote{http://www.lofar.org}, 
LWA\footnote{http://lwa.phys.unm.edu}, 
MWA\footnote{http://www.mwatelescope.org}, 
PAPER\footnote{http://eor.berkeley.edu}, etc. 
In 2004, we began to construct a unique 
low frequency radio array, the 21 CentiMeter Array (21CMA), 
with the goal of performing a statistical measurement of the power spectrum 
of EoR. A total of 10 287 log-periodic antennas were deployed along two 
perpendicular arms with lengths 6 and 4 km, respectively. We completed 
the construction in 2007 and upgraded the system in 2010. Sited in the 
Tianshan Mountains, west China, the array has been collecting data towards 
the north celestial pole (NCP) region for nearly five years. 

One of the major difficulties in current SKA pathfinders or precursors 
towards the detection of the cosmic dawn and EoR is that the cosmic signal 
is only 1-10 mK, but strong RFI may still be present at low frequencies 
even if radio telescopes are placed in remote areas. 
This is because there exist two types of man-made celestial sources of RFI:  
(1) FM radio and TV broadcasting scattered by meteor trails and/or aircraft
and (2) satellite communications, in addition to system noise. 
In addition, the radio foreground dominated primarily 
by the Milky Way is roughly 4-5 orders of 
magnitude brighter than the signal to be measured. How to mitigate all 
these RFI and suppress the foregrounds to a desirable level really 
poses a technical challenge for on-going and planned experiments. 
If man-made, time-varying RFI and system noise can be well understood 
and perfectly excised, there are essentially four steps towards 
the removals of bright foreground for a radio interferometer:  
First is to reduce the thermal noise through interferometric 
correlation, and then integrate for a sufficiently long time even to the 
confusion limit. Second is to identify and subtract all the bright 
point sources including those from the leakage of sidelobes, 
through either the conventional Cotton-Schwab CLEAN algorithm 
(Schwab ~\cite{Schwab84}) or a more advanced approach 
such as ``Peeling'' (Noordam ~\cite{Noordam82}). 
Third is to remove a power-law or smooth component 
in the frequency domain because the low frequency foreground is dominated by 
synchrotron radiation which is believed to be spatially and spectrally smooth.  
It has been shown by recent simulations (e.g. Alonso et al. ~\cite{Alonso15})   
that even the blind method without the assumption of a foreground 
spectrum can yield similar and satisfactory results. 
Finally, instead of the EoR imaging which will be achieved by a next 
generation radio interferometer like SKA, 
we can statistically construct the power spectrum of the low frequency sky, 
which allows us to further ``beat down'' the sky noise by a factor 
of $1/N_{\ell}$, where $N_{\ell}$ represents the independent Fourier modes 
of the targeted field.  

Detection and mitigation of RFI is the first and also a key step for 
current low frequency radio experiments like 21CMA, 
and the quality and accuracy of the RFI excision 
critically determines whether our science goals can be 
eventually achieved. In this work, we present the radio environment report 
of the 21CMA, demonstrating the various techniques for 
identification and removals of RFI in the 21CMA observations. 
Similar radio environment reports on LOFAR, MWA and LWA can be found 
in Offringa et al. (~\cite{Offringa13}); Offringa et al. (~\cite{Offringa15}) 
and Henning et al. (~\cite{Henning10}), respectively. 
Besides the traditional approach, we mainly concentrate on 
the statistical analysis of RFI detection, in the sense that both 
system noise, including sky noise and cosmological signal like EoR 
behave like thermal noise at low frequencies, and therefore should 
statistically follow a Gaussian distribution. If we mitigate 
all the non-Gaussian components, most of the temporal RFI above 
the thermal noise should be able to be excised. 
This can be made either in real time or in 
data post processing, and the accuracy 
depends sensitively on time and frequency resolution. 
Moreover, we introduce the so-call visibility correlation 
coefficient instead of the conventional visibility, allowing a further 
reduction of the RFI amplitude. Finally, we briefly discuss the 
removal of the sky RFI introduced by the leakage of sidelobes from 
off-field strong radio sources.

\section{System noise: RFI identification}
\label{sect: RFI}

Radio telescopes, when working at low frequencies of $\nu\le200$ MHz, are 
primarily limited by spatial resolution. For example, their effective 
apertures should be as large as 10 km to achieve an angular resolution of 
$\sim1'$. Employment of radio interferometry turns out to be a unique 
solution to this problem for low frequency observations. To be specific, 
signal received by the $i-$th antenna unit in an interferometric 
array at time $t$ is usually represented by the voltage $V_i$ 
\begin{equation}
V_i(t)=G_i\int B_i(\mbox{\boldmath s})E(\mbox{\boldmath s},t)
         d^2\mbox{\boldmath s} + \epsilon_i,
\end{equation} 
where $G_i$ is the system gain, $ B_i(\mbox{\boldmath s})$ 
is the spatial response function (or sometimes primary beam) of 
the antenna, $E(\mbox{\boldmath s},t)$ is the electric field of the 
cosmic signal along direction $\mbox{\boldmath s}$ and can also be 
regarded as the Fourier component for a given frequency $\nu$, 
$\epsilon_i$ is the system noise, and the integral should be taken over 
all directions being observed. Note that all these quantities are a 
function of $\nu$.  
The total power that the $i-$th antenna receives is given by 
the autocorrelation of Eq.(1)
\begin{equation}
V_{ii}(t)=\left|G_i(t)\right|\int B_{ii}(\mbox{\boldmath s})
         \left|E(\mbox{\boldmath s},t)\right|^2
         d^2\mbox{\boldmath s} + \left|\epsilon_i\right|^2,
\end{equation} 
while the cross-correlation of Eq.(1) yields the so-called visibility 
function 
\begin{equation}
V_{ij}(t)=G_i(t)G^*_j(t) \int B_{i}(\mbox{\boldmath s})
                            B_{j}^*(\mbox{\boldmath s})
         \left|E(\mbox{\boldmath s},t)\right|^2 e^{i2\pi\nu\Delta t}
         d^2\mbox{\boldmath s} + \epsilon_i\epsilon_j^*,
\end{equation} 
in which the raised asterisk indicates the complex conjugate, 
$\Delta t$ is the time delay of received signal between 
the $i-$th antenna and $j-$th 
antenna due to geometric configuration, and the coherence condition of 
astronomical radiation has already been used. Because noise in two antenna 
systems is not coherent, the time averaged operation of 
$\langle \epsilon_i\epsilon_j^*\rangle$ will significantly reduce 
the noise level, and sometimes the last term in Eq.(3) can even be neglected. 
Quantitatively,   $\langle \epsilon_i\epsilon_j^*\rangle$ is 
$1/\sqrt{\Delta\tau\Delta t}$ times smaller 
than the noise power of a single antenna  $\left|\epsilon_i\right|^2$ or 
$\left|\epsilon_j\right|^2$, where 
$\Delta \tau$ is integration time and $\Delta\nu$ is bandwidth.

Fig.1 shows a typical example of the total powers of 
auto- and cross-correlations. Data are taken from 24-hour observations 
on 2013 May 11 for two antenna pods E01 and E20 
along the east-west baseline of 21CMA, separated by 1280 m. 
(The same data set will be used below unless otherwise stated.) 
The signal is digitized at a sampling rate of 400 MHz with 8 bit precision. 
We adopt a frequency channel of 8192, which provides a resolution 
of 24.2 kHz over a bandwidth of 200 MHz. 
Fast Fourier transform (FFT) and correlation calculations are performed 
in real time through software, and the digitized and correlated data 
are integrated for about 3.558 seconds before they are output to a hard disk. 
A total of 24 280 visibilities including autocorrelations are recorded 
in each channel for offline analysis. From the average power displayed 
in Fig.1, we learn that the power or autocorrelation of a single antenna 
pod is dominated by the system thermal noise $\left|\epsilon_i\right|^2$ or 
$\left|\epsilon_j\right|^2$, and $V_{ii}$ and  $V_{jj}$ are almost three orders 
of magnitude larger than the power of visibility $|V_{ij}|$. As mentioned 
above, thermal noise follows a  Gaussian distribution and  
$\langle \epsilon_i\epsilon_j^*\rangle$ is therefore reduced by a factor of 
$1/\sqrt{N}$ as compared with $\left|\epsilon_i\right|^2$ or 
$\left|\epsilon_j\right|^2$, 
where $N$ is the total number of samples over an integration time 
of $\Delta \tau$ and a bandwidth of $\Delta\nu$.  This yields 
$1/\sqrt{N}=1/\sqrt{\Delta\tau\Delta\nu}=
1/\sqrt{24kHz\times3.558s}=3.4\times10^{-3}$. The actual noise level 
of $|V_{ij}|$ shown in Fig.1 is slightly higher than this simple estimate 
because the efficiency of the 21CMA ($50\%$) has not been taken into account.  

\begin{figure}
\centering
\includegraphics[width=10cm, angle=0]{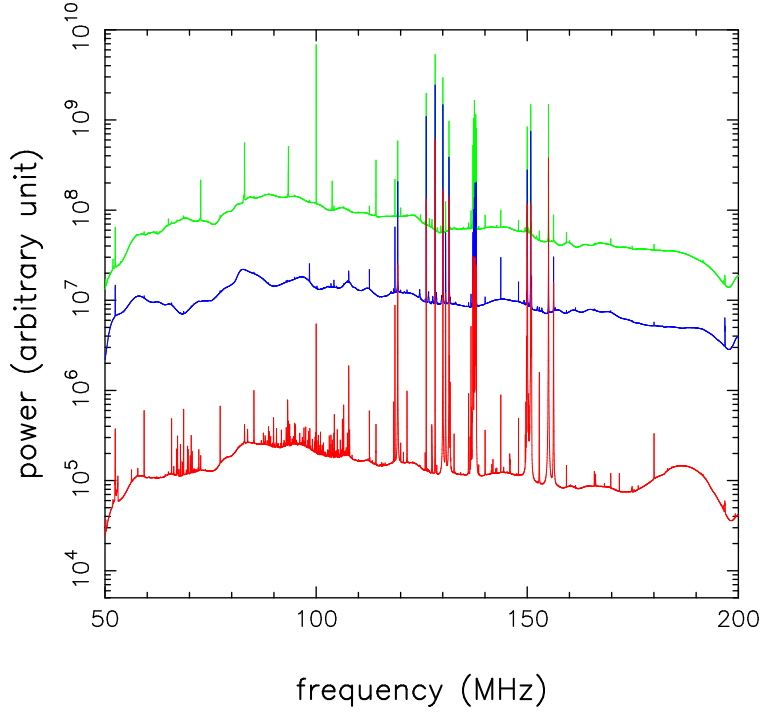}
 \caption{Average powers of autocorrelation for E01 (blue) and 
E20 (green) and their cross-correlation (red).} 
\label{Fig:plot1}
\end{figure}

Because the system temperature is dominated by thermal noise 
(the receiver noise of 21CMA is about 50 K), 
only a few strong RFI features above the thermal noise level are present 
in the autocorrelation spectrum, as shown in Fig.1. However, after the thermal 
noise is suppressed by three orders of magnitude, the power of visibility 
seems very noisy, and a glimpse of the spectrum in Fig.1 reveals many strong 
RFI signatures. Yet, most of them actually arise from time-varying RFI and 
sparks, and can be easily identified and mitigated. The 
strong RFI features in $|V_{ij}|$ are identified as follows: 
(1) Civil aviation communications 
at 119 MHz and 130 MHz, produced by commercial aircraft which last 4 minutes 
each time when an aircraft flies over the 21CMA valley. 
(2) Low earth orbit satellite broadcasting at 137 MHz, generated 
by the ORBCOMM constellation. The good sky coverage of the ORBCOMM satellites 
indicates that they are the most prominent RFI source 
for all low frequency 
radio telescopes even at remote sites. Furthermore, since there are always 
several ORBCOMM satellites above the horizon at any given time, 
the 137-138 frequency band can hardly be used. Yet, they may 
serve as a beacon for calibration of our telescopes (e.g. Neben et al. 
~\cite{Neben15}). 
(3) Walkies-talkies from local train communications 
at 151 MHz, which often cause the problem of saturation when trains pass 
through the 21CMA site. As a result, some of the data in the 150-151 band 
have to be flagged for blanking. 
Fortunately, the old railroad track at the site is no longer 
used and the strongest interference at 151 MHz has thus 
disappeared since January 2014. 
(4) FM radio broadcasting at 88-108 MHz, which is scattered by meteor 
trails and aircraft. Note that the site, surrounded by mountains with 
altitudes over 3000 m, cannot receive FM transmissions from local towns. 
(5) AM radio broadcast around 70 MHz, which is time variable and rather weak. 
(6) Computer noise at 100 MHz and 107.7 MHz, which appears occasionally and 
is sharp but very strong. For the sample (2013 May 11, observation) 
we chose in this paper, the computer RFI at two narrow bands of 100 MHz 
and 107.7 MHz remains active through the whole observing time. The source 
of noise is identified as being due to the synchronization clock used in 
our other experiment on radio detection of cosmic air showers and 
cosmic neutrinos (Ardouin et al. ~\cite{Ardouin11}).

\section{System noise: statistical treatment}
\label{sect:statistics}

Most of the RFI shown in Fig.1 is highly varying in time and can be easily 
detected and mitigated, though the accuracy and efficiency depend 
on time and frequency resolutions. For example, setting a threshold 
in visibility may allow us to excise most of the strong RFI. 
Alternatively, one may use a statistical or 
automated method to identify RFI in visibility. Note that 
both system noise (including sky noise) and cosmic radio 
sources behave like thermal noise, and therefore the visibility data 
without terrestrial RFI should exhibit a Gaussian distribution. 
Now, if we identify and remove all the non-Gaussian structures in 
each channel, RFI should be effectively mitigated. Such an approach 
can be applied to either data acquisition in real time, if a proper trigger 
(or standard deviation) is set for each channel, or data post processing. 
The latter seems more flexible and convenient for users 
but could result in inaccuracy of RFI detection due to the lack of high time 
resolution. To guarantee the efficiency of the 21CMA data acquisition which 
uses software to do FFT and correlations, here we demonstrate the statistical 
mitigation of RFI only during data post processing. 

Fig.2 shows the time variation of the visibility power $|V_{E01W20}|$ for 
antenna pair E01E20 at a central frequency of 160 MHz and with a bandwidth 
of 0.1 MHz. If we directly apply Gaussian statistics to the data, we 
probably need to deal with the problem of large dynamical range. 
Meanwhile, we also lose the phase information in visibility. 
Therefore, we begin by normalizing 
the real and imaginary parts of visibility separately, 
i.e. dividing by their own mean amplitudes, so that the real and imaginary 
parts of visibility are both varying around `0'. In Fig.3 we illustrate 
the normalized real part of the visibility $V_{E01W20}$ versus time. If we 
excise the data points beyond $5\sigma$, all the strong RFI can be 
mitigated. Setting a tight limit of $3\sigma$ should allow us 
to exclude most of the RFI yet at the cost of losing or distorting some 
of the true signal. The choice of the excision limit varies among different 
experiments. For example, PAPER adopts a limit of $6\sigma$ in 
Pober et al. (~\cite{Pober13}) and $3\sigma$ 
in Parsons et al. (~\cite{Parsons14}). 
When processing the 21CMA data, we work with 
an adaptive choice of the $\sigma$ range. For the channels that are 
frequently contaminated, such as 121 MHz, 137 MHz (see Fig.4), and 151 MHz, 
we adopt a limit of $3\sigma$. Otherwise, we use a more relaxed choice of 
$5\sigma$ (see Fig.3) because the majority of the channels actually remain 
very quiet.  For comparison in Fig.4, we provide an extreme example of the 
most seriously contaminated case at 137 MHz along with 
the $3\sigma$ and $7\sigma$ limits.  

With the above RFI detection limit, we can now calculate the 
RFI occupancy in different frequencies, which provides a quantitative 
description of the RFI influence on our radio observations. In Fig.5 
we display the RFI occupancy spectrum for E01E20. The most contaminated 
channels are found to be 100 MHz and 107.7 MHz - the sharp noise from 
the synchronization clock used in our experiment on cosmic rays at the same 
site, followed by the ORBCOMM satellite signal around 137 MHz with a 
maximum occupancy of $33\%$. 
Some of the channels below $\sim80$ MHz seem a bit noisy and only $\sim90\%$ 
of the time can be used for observations. Local train communications 
at 150 MHz lead to $13\%$ of the data that should be excised. 
Also, there is a slowly varying occupancy ratio of $\sim 2-6\%$, 
spanning a wide range of frequencies from 175 MHz to 200 MHz, 
which is attributed to the active radio emission region from 
our Galaxy and will be addressed extensively in Section 5. 
For comparison, LOFAR and MWA have provided their occupancy 
ratios in a sub-range of 115-163 MHz, which are $3.18\%$ and $1.65\%$, 
respectively (Offringa et al. ~\cite{Offringa13}; 
Offringa et al. ~\cite{Offringa15}). 
Our result in the same frequency range is $2.66\%$. 
 
\begin{figure}
\centering
\includegraphics[width=10cm, angle=0]{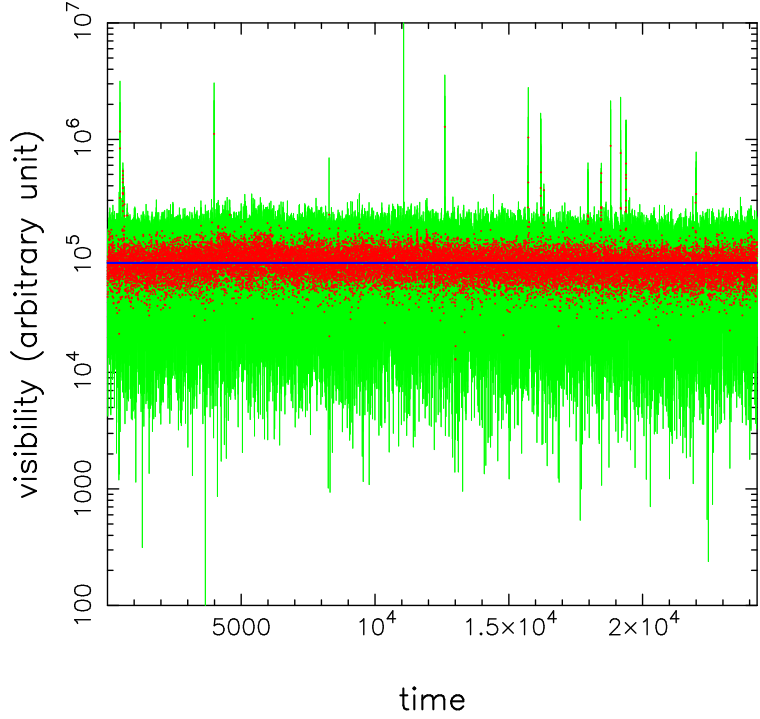}
 \caption{Time variation of visibility power for the antenna pair E01E20 at 
a central frequency of 160 MHz with bandwidth 0.1 MHz (four channels). Red 
points represent the instantaneous average among four channels, while a blue 
line is the average over an integration time of 24 hours.} 
\end{figure}

\begin{figure}
\centering
\includegraphics[width=10cm, angle=0]{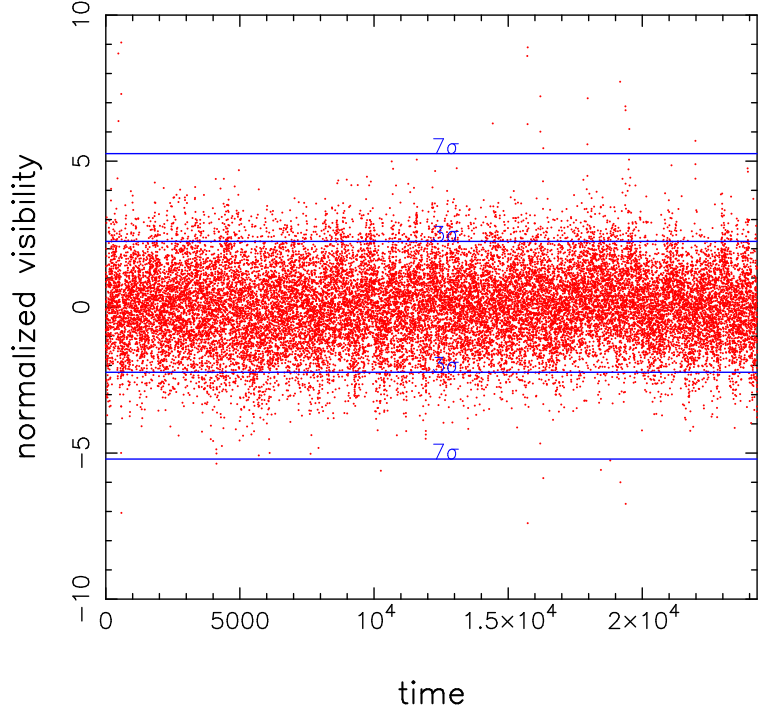}
 \caption{The real part of the normalized visibility $V_{E01W20}$ 
versus observing time at frequency $\nu=160$ MHz.} 
\end{figure}

\begin{figure}
\centering
\includegraphics[width=10cm, angle=0]{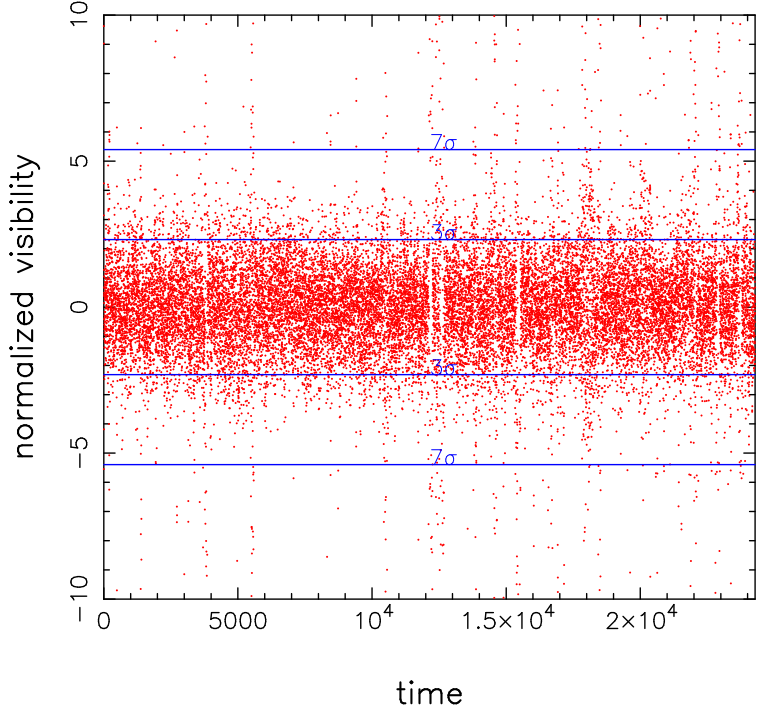}
 \caption{The same as Fig.3 but for $\nu=137$ MHz.} 
\end{figure}

\begin{figure}
\centering
\includegraphics[width=10cm, angle=0]{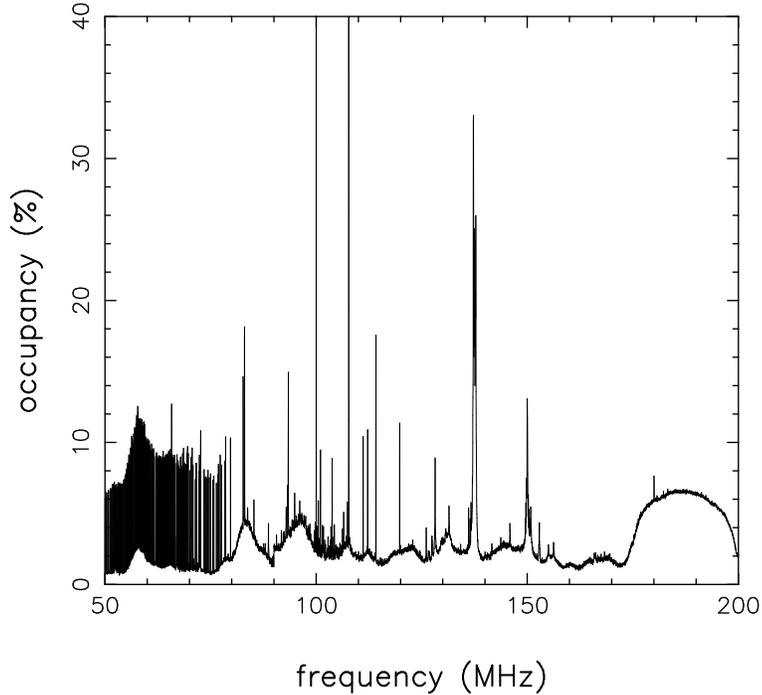}
 \caption{RFI occupancy spectrum for E01E20, in which RFI is selected 
in terms of $5\sigma$.} 
\end{figure}

Fig.6 shows the RFI mitigated power of the visibility $V_{E01W20}$ 
in terms of the non-Gaussian criterion. As compared with Fig.1, 
most of the RFI has been excised or significantly suppressed. 
It is necessary to make a few remarks on this RFI mitigated visibility 
power: The overall contour is a convolution of the antenna spectral 
response with the low frequency radio sky dominated by the Milky Way. 
The decreasing power with increasing frequency in 85-200 MHz just represents 
the Galactic radiation, while the turnover near 85 MHz reflects the 
inefficient antenna gain at lower frequencies. Actually, the design of 
the 21CMA log-periodic antenna is only optimized for 70-350 MHz. Finally, 
the rapid attenuation at the two frequency ends arises from the roll-off 
of the two bandpass filters used in our receiver. In principle, the data 
from adjacent channels should be excluded. 

\begin{figure}
\centering
\includegraphics[width=10cm, angle=0]{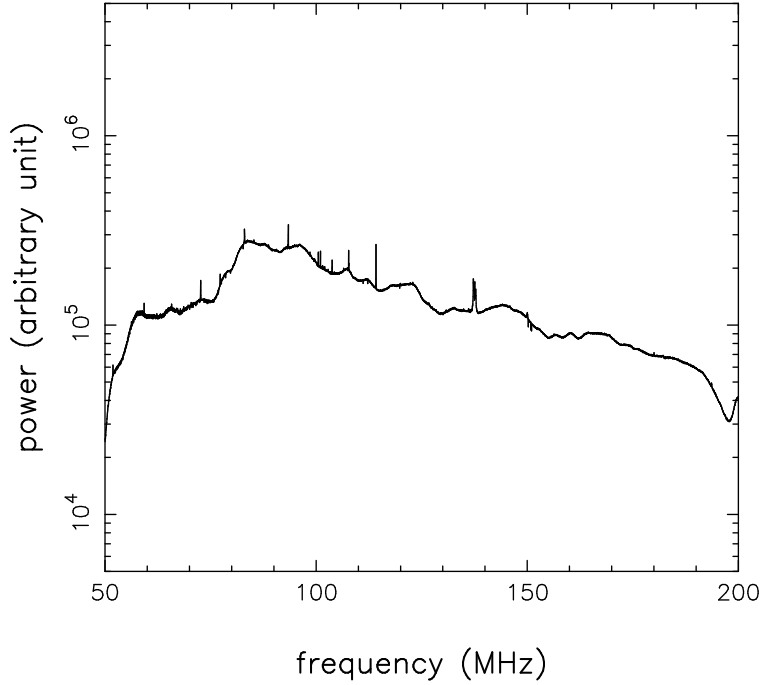}
 \caption{Average power of the visibility $V_{E01W20}$ after mitigating 
the non-Gaussian features. See Fig.1 for the original data.} 
\end{figure}

Even after the temporal RFI is mitigated, system noise still 
dominates the visibility function at low frequencies, 
and a long integration is required to ``beat down'' noise that 
mainly comes from the Milky Way. For experiments toward the EoR detection,  
integrating observations may take up to a few years. Therefore, it is 
crucial to check whether and how the system noise goes down 
with integration time. Telescope sensitivity in terms of flux can be written 
as $2k_BT_{\rm sys}/A_{\rm eff}\sqrt{t\Delta\nu}$, where $k_B$ is the 
Boltzmann constant, $T_{\rm sys}$ is the system temperature, and $A_{\rm eff}$ 
is the effective area of the telescope. In our case, all antenna pods 
have the same effective area, and their receivers have roughly the same 
system temperature. It turns out that theoretically, 
for a given frequency bandwidth, 
the sensitivity should vary with integration time as $t^{-1/2}$. 
In Fig.7 we plot the average visibility power $|V_{E01W20}|$ and its variance 
at $\nu=160$ MHz against integration time for observations spanning over 
24 hours. It appears that the sensitivity does follow the $t^{-1/2}$ 
behavior. Similar tests have also been made for other EoR experiments such 
LWA, MWA, etc. (Bowman et al. ~\cite{Bowman07}; 
Henning et al. ~\cite{Henning10}; Taylor et al. ~\cite{Taylor12}; 
Offringa et al. ~\cite{Offringa15}).  

\begin{figure}
\centering
\includegraphics[width=10cm, angle=0]{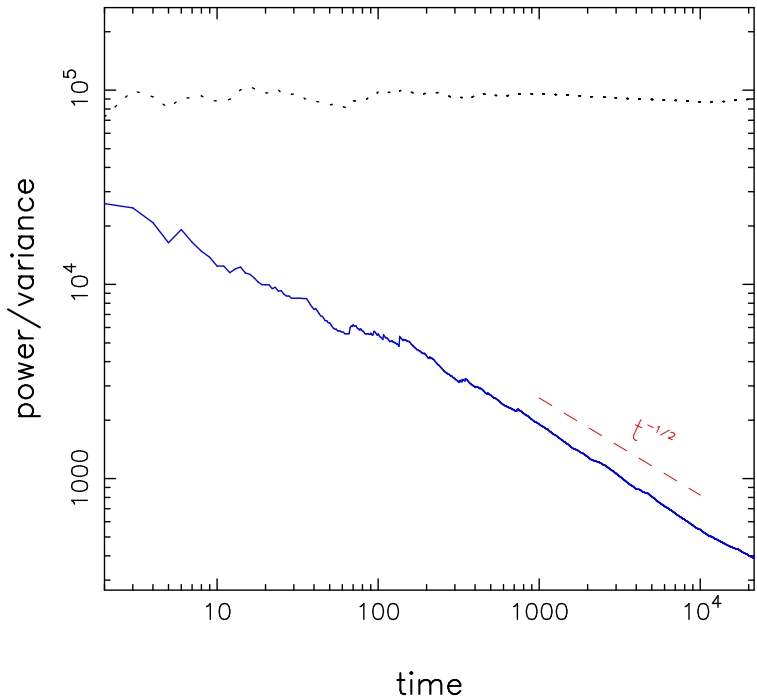}
 \caption{Average visibility power $|V_{E01W20}|$ (black) and its 
variance (blue) at $\nu=160$ MHz versus integration time. 
The latter continues to drop as $t^{-1/2}$ over 24 hours.}
\end{figure}

\section{Visibility correlation coefficient}
\label{sect:correlation}

We may take a step further to introduce the so-called visibility 
correlation coefficient in the analysis of RFI mitigation. 
The visibility correlation coefficient is defined as 
\begin{equation}
p_{ij}(t)=\frac{V_{ij}(t)}{\sqrt{|V_{ii}(t)||V_{jj}(t)|}}.
\end{equation} 
We begin with two extreme cases to demonstrate the advantage of utilizing 
$p_{ij}$ instead of $V_{ij}$: (1) If the visibility and 
autocorrelations are dominated by RFI, then the above expression reduces
to $p_{ij}(t)=\epsilon_{ij}(t)/\sqrt{|\epsilon_{ii}(t)||\epsilon_{jj}(t)|}$. 
All strong RFI in antenna unit $i$ and/or $j$ can cancel each other through
the $|p_{ij}(t)|$ operation. 
Namely, $p_{ij}$ can suppress the amplitude of strong RFI as it appears 
simultaneously in the numerator and denominator. 
(2) If the noise terms $\epsilon_{ii}$,  $\epsilon_{jj}$ and  $\epsilon_{ij}$ 
are negligibly small, then the power of $p_{ij}$ can be converted into 
the angular power spectrum $C_{\ell}$ through 
$C_{\ell}=I_0^2|p_{ij}(\ell/2\pi)|^2$ 
(Zheng et al. ~\cite{Zheng12}), in which $I_0$ is 
the mean brightness of the observed radio sky. In other words, $p_{ij}$ 
provides a simple way to statistically measure the angular power spectrum.

Yet in reality, system thermal noise $\epsilon_{ij}$ including the 
Milky Way contribution is comparable to the total flux of cosmic radio 
sources at frequencies below 200 MHz. Namely,
$\langle 
G_iG^*_j \int B_{i}(\mbox{\boldmath s})B_{j}^*(\mbox{\boldmath s})
 \left|E(\mbox{\boldmath s})\right|^2 e^{i2\pi\nu\Delta t}
         d^2\mbox{\boldmath s} \rangle$
and $\langle \epsilon_i\epsilon_j^*\rangle$ have the same order of magnitude, 
unless there is strong RFI in $\epsilon_{ij}$. 
Indeed, this is supported by the following argument: 
The Galactic radiation is the main source of noise in current EoR experiments 
if the receiver noise is controlled within $\sim50$ K. The surface 
temperature of the Milky Way can be estimated 
by $T_b=200 {\rm K}(\nu/150{\rm MHz})^{-2.5}$. 
Its rms variation through the $\langle \epsilon_i\epsilon_j^*\rangle$ 
operation is modulated by a factor of $1/\sqrt{\Delta\nu\Delta\tau}$: 
$\Delta T_b=T_b/\sqrt{\Delta\nu\Delta\tau}$. For the EoR experiment like 
21CMA, $\Delta T_b$ is approximately a factor of 1000 smaller than $T_b$. 
Converting  $\Delta T_b$ into the rms variation in flux yields 
\begin{equation}
\Delta S=0.9{\rm Jy}\left(\frac{\lambda}{3{\rm m}}\right)^{-2}
            \left( \frac{\Delta T_b}{0.1{\rm K}}\right)
            \left( \frac{\theta_b}{10^{\circ}}\right)^2,
\end{equation}  
where $\theta_b$ is the width of the telescope primary beam. This simple 
estimate indicates that the system noise is roughly on the same order 
of magnitude as the bright radio sources (0.1 Jy - 10 Jy) in the sky. 
Only when RFI is brighter than the sky noise  $\Delta T_b$, i.e., the RFI 
plays a dominant role in $\langle \epsilon_i\epsilon_j^*\rangle$ and is 
also greater than 
$\langle 
G_iG^*_j \int B_{i}(\mbox{\boldmath s})B_{j}^*(\mbox{\boldmath s})
 \left|E(\mbox{\boldmath s})\right|^2 e^{i2\pi\nu\Delta t}
         d^2\mbox{\boldmath s} \rangle$, 
does the employment of $p_{ij}$ become useful and efficient in 
mitigation of RFI. Although it may be rather difficult to identify weak 
RFI that is buried under sky noise, it is always helpful to 
utilize the visibility correlation coefficient for an efficient 
detection and mitigation of strong RFI. 

Now we demonstrate the effect and result of RFI mitigation in terms 
of $p_{ij}$ for the 21CMA data. There are two ways to employ $p_{ij}$ in 
data processing: First is to compute $p_{ij}$ from visibility 
$V_{ij}$ and antenna powers $V_{ii}$ and $V_{jj}$ in data post processing, and 
work with $p_{ij}$ instead of $V_{ij}$ in subsequent image processing. 
Second is to add a new task of computing and integrating $p_{ij}$ 
to the data acquisition phase in real time. The former does not 
add any extra work in the data acquisition system, but the  
RFI detection may be affected by poor time resolution in 
output visibility for which integration has already been done. The latter 
can maximally guarantee the accuracy of RFI identification and mitigation, 
at the cost of decreasing the efficiency of data acquisition, because more 
CPU/GPU time is required to compute $p_{ij}$ and even complete the 
real-time RFI detection before integration and output operations. 
Leaving the first approach to data post processing, 
we tested the second approach with 21CMA during September, 2012. A total 
of 26 days were devoted to the measurement of $p_{ij}$ instead of $V_{ij}$, 
for which the telescope efficiency is found to further decrease by $22\%$, 
in addition to the $50\%$ efficiency in its routine observing mode. 
In Fig.8 we compare the visibility power  $|V_{ij}|$, correlation coefficient 
$|p_{ij}|$ in data post processing and $|p_{ij}|$ at real time, in which 
the first two results are based on the 2013 May 11 observation, and the 
third one is taken from the 2012 September 19 measurement. 
In both case, the same baseline E01E20 is chosen. 
We take a frequency channel at $\nu=102$ MHz with a 0.1 MHz 
bandwidth for the purpose of illustration, because the channel is 
occasionally contaminated by FM radio signals scattered by 
meteor trails and/or aircraft. A visual inspection of Fig.8 reveals 
that the amplitudes of strong RFI in $|p_{ij}|$ become 
smaller than those in $|V_{ij}|$. To quantitatively evaluate the effect of 
RFI detection from the three methods, we calculate their mean values 
$\bar{V}$ or $\bar{p}$ and variances $\sigma$. Furthermore, we utilize 
$\sigma/\bar{V}$ or $\sigma/\bar{p}$ to characterize the extent of their 
dispersion and show the results in Fig.9. It appears that the visibility 
function suffers from the most serious contamination as indicated by 
the largest dispersion in $\sigma/\bar{V}$, though most of these RFI are 
actually time-varying or sparks. Note that the same data are displayed in 
a different way in Fig.1. On the contrary, $|p_{ij}|$ 
can significantly suppress the effect of RFI, as manifested by 
much smaller dispersion in $\sigma/\bar{p}$. It seems that the real-time 
operation of $|p_{ij}|$ yields an even better result. 
If we take into account the fact that the real-time computation of $p_{ij}$ 
reduces the efficiency of data acquisition to some extent, the data 
quality of $\sigma/\bar{p}$ can be further improved with the employment 
of high-speed computers or field-programmable gate arrays (FPGAs) for FFT 
and correlations.

\begin{figure}
\centering
\includegraphics[width=10cm, angle=0]{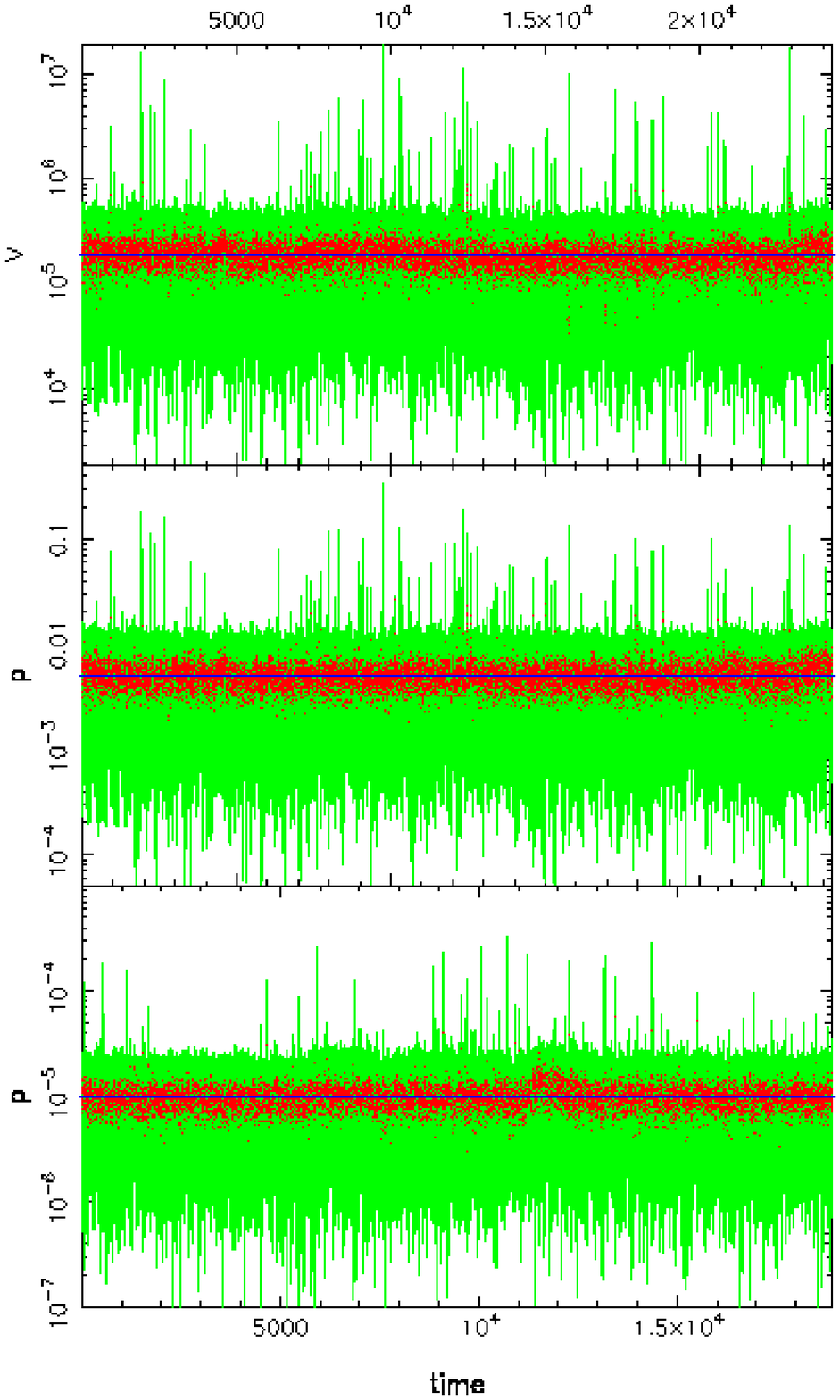}
 \caption{Visibility power  $|V_{E01E20}|$ (top), 
visibility correlation coefficient power $|p_{E01E20}|$ with 
data post processing approach (middle), and $|p_{E01E20}|$ 
with real-time operation (bottom). The central frequency is 102 MHz with 
a bandwidth of 0.1 MHz. The average among different channels at a given 
timestep is represented by red points, and the average over the whole 
integration time is indicated by a blue line.}
\end{figure}

\begin{figure}
\centering
\includegraphics[width=10cm, angle=0]{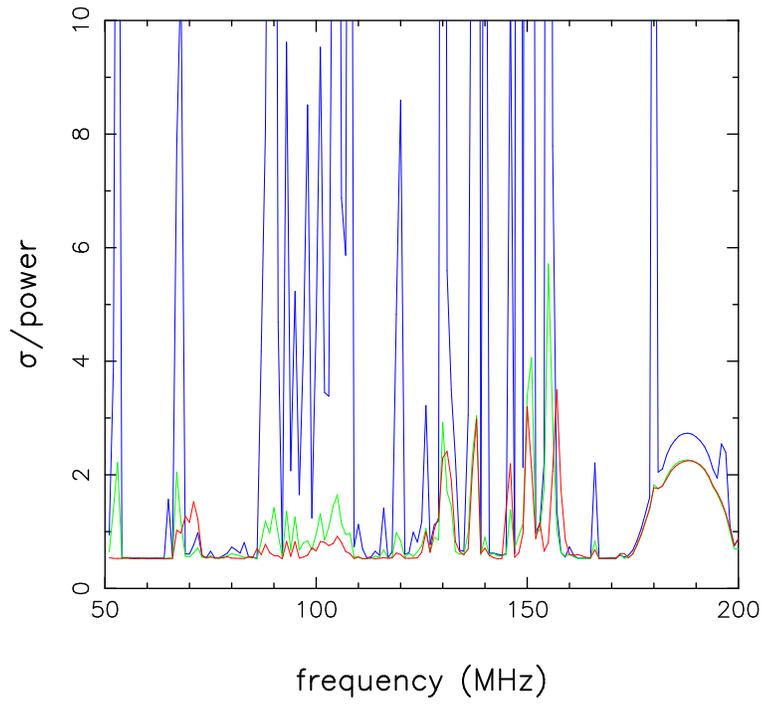}
 \caption{Comparisons of data dispersion, represented by the ratio of 
standard deviation to mean value for three different approaches. 
The blue line represents $V_{E01E20}$; the green line represents $p_{E01E20}$ 
with data post processing; and the red line represents 
real-time $p_{E01E20}$.}
\end{figure}

\section{RFI in the Sky}
\label{sect:RFI sky}

Unlike the man-made, terrestrial RFI that we have discussed above, 
celestial RFI is actually emitted by strong radio sources in the sky. 
But for the EoR experiments at low frequencies, they can also be treated 
as a sort of RFI and should be mitigated. The dominant radio source in the 
low frequency sky is of course the Milky Way, and its global, 
uniform radiation has already been included in the noise term 
$\epsilon$ above. Now, what we will deal with is the structured components 
that can be captured by radio interferometers, including both Galactic 
and extragalactic strong radio sources. 

Fig.10 is the waterfall plot for the real part of visibility 
$V_{E01E20}$ observed on 2013 May 11, which illustrates the 24 hour 
variation in the radio environment containing both terrestrial RFI and 
celestial sources. Three predominant features from the sky are identified 
as follows: 
(1) Thick fringes; these correspond to the brightest radio source (quasar) 
within 1 degree of NCP, B004713+891245, with a flux of 5 Jy and flat 
spectral index. 
(2) Thin fringes; these are generated by the brightest radio galaxy, 
3C061.1, located $\sim4$ degrees from NCP. Extrapolating its spectral index 
of $-0.8$ to 150 MHz, we find that it has a flux of roughly 33 Jy. 
(3) Two bright spots with noticeable fringes and an interval of 6 hours, 
which are the strongest RFI source in the 21CMA NCP field and will be the 
main focus in this section. Actually, we have already seen their effect 
through the bumps at $\nu=175-200$ MHz shown in Figs.1, 5 and 9. 
For the two bright extragalactic sources, B004713+891245 
and 3C061.1, we will include them in the source catalog observed with 
21CMA and present their properties and excision elsewhere 
(Zheng et al. ~\cite{Zheng16}). 

\begin{figure}
\centering
\includegraphics[width=14cm, angle=0]{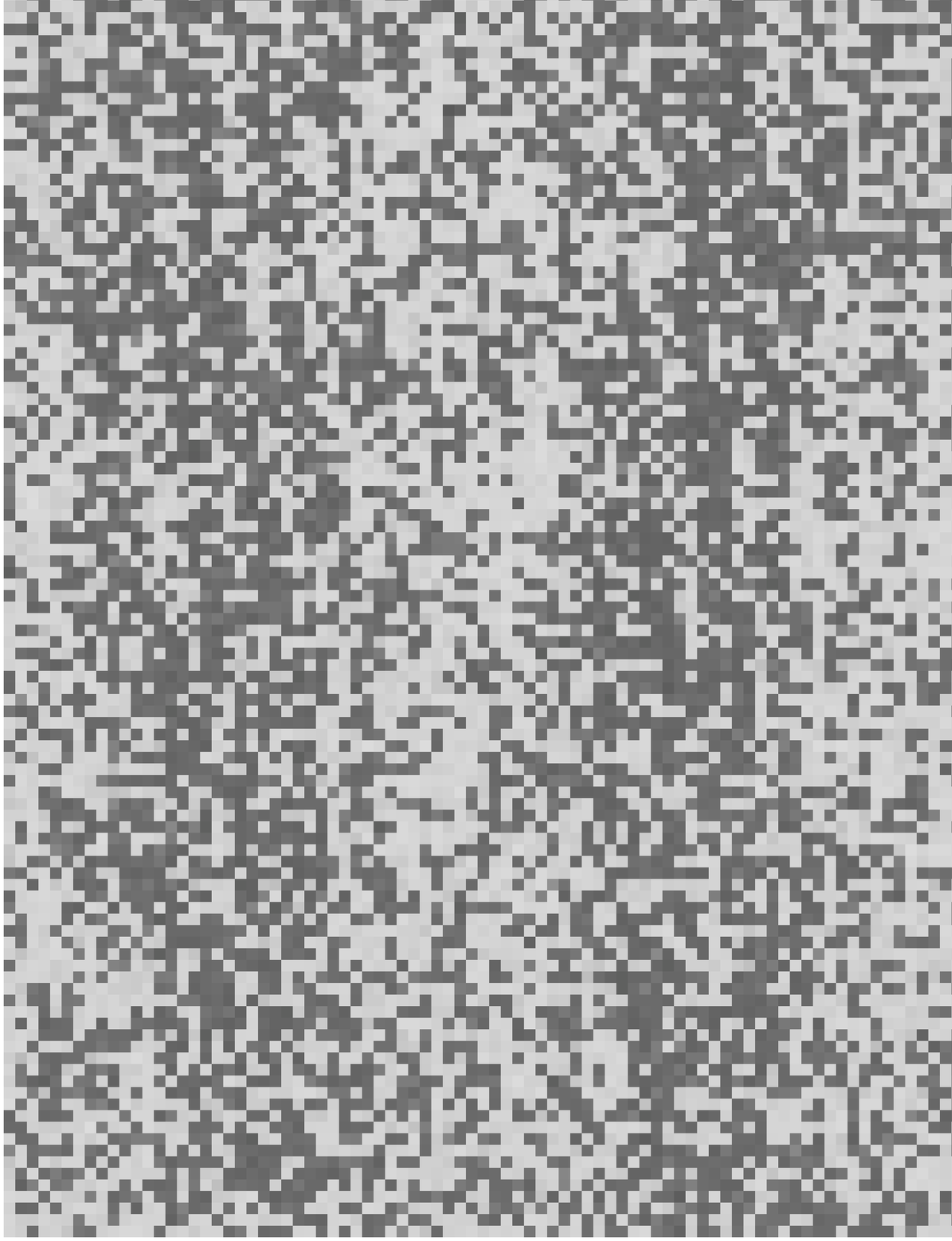}
 \caption{The waterfall plot of the real part of visibility function 
$V_{E01E20}$ spanning over 24 hours along the horizonal axis. The vertical 
axis represents frequency running from 50 MHz (top) to 200 MHz (bottom).}
\end{figure}

In order to understand the origin of the two strong RFI features around 
180 MHz with interval time of 6 hours, we begin with the beam pattern of 
the 21CMA pod. Each of the 21CMA pods consists of 127 log-periodic antennas 
which are combined though a phase delay cable to form an analog beam towards 
NCP. The primary beam, centered on NCP, has a field of view of approximately 
$8.5^{\circ}(\nu/100{\rm MHz})^{-1}$. However, the regular spacing of 
127 antennas in the hexagonal pod also leads to two remarkable sidelobes, 
separeted by 90 degrees with respect to the NCP. Their radii vary with 
frequency, and reach about $45^{\circ}$ near $\nu=175-190$ MHz.   
In Fig.11(a) and Fig.12(a) we display the beam patterns of 21CMA for 
$\nu=150$ MHz and  $\nu=180$ MHz, respectively. If no 
radio sources fall into the sidelobes during diurnal motion of the 
heavens, then our observation towards the NCP field is unaffected by 
existence of the sidelobes. Otherwise, the sidelobe leakage of off-field 
radio sources may contaminate our observation and should be removed. 
Unfortunately, as the dominant source of radio emission at low-frequencies, 
the Galactic plane sweeps the NCP region every 24 hours, which brings 
the Galactic emission into the 21CMA field of view twice per day through 
the two sidelobes. In particular, a large star forming region called Cygnus A 
is just located about $45^{\circ}$ from NCP, and can be clearly seen 
through the two sidelobes of the 21CMA at frequencies around 180 MHz. This 
gives rise to the two bright spots, separated by exactly 6 hours, in the 
waterfall plot shown in Fig.10. These regions that cause contamination 
will be a disaster for our interferometric imaging of the EoR sky and should 
certainly be excised. On the other hand, the sidelobe leakage may provide 
a way to study the large star forming region in Cygnus, e.g., by monitoring 
of time variation using long integration data of the EOR spanning years.

\begin{figure}
\centering
\includegraphics[width=10cm, angle=0]{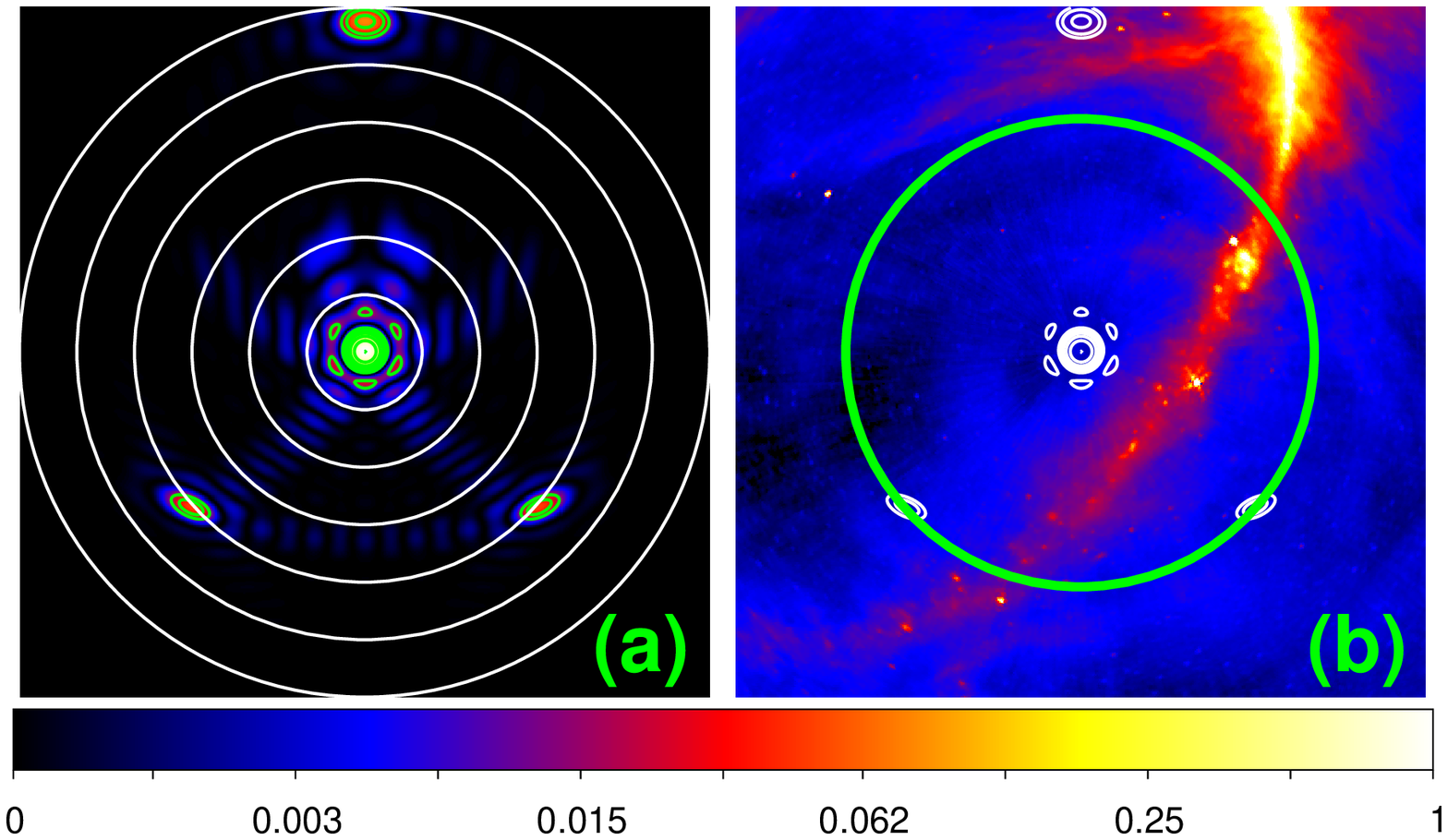}
 \caption{The beam pattern of the 21CMA at $\nu=150$ MHz (a) overlaid on 
the sky map of Haslam et al. ~\cite{Haslam82} for 408 MHz (b). 
An annulus with a diameter of 10 degrees indicates the area shown (a). 
The green circle in (b) indicates the locus of diurnal motion of the 
Galactic region that passes through the sidelobes of the 21CMA. }
\end{figure}

\begin{figure}
\centering
\includegraphics[width=10cm, angle=0]{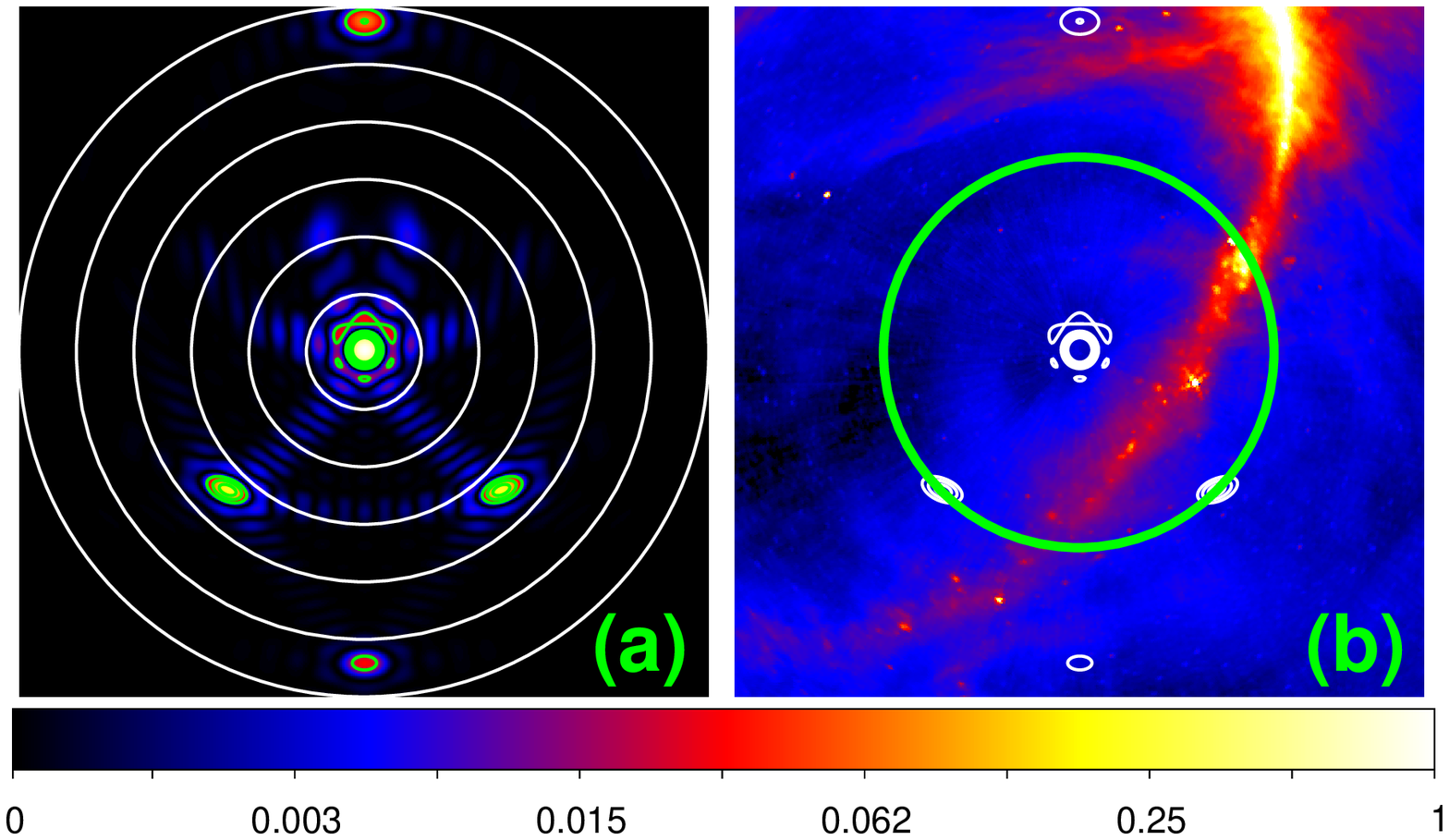}
 \caption{The same as Fig.11 but for $\nu=180$ MHz. The diurnal motion of 
the Cygnus region traces a circle with a radius of $\sim45^{\circ}$ from NCP, 
which enables the sidelobes to `see' the Cygnus region twice per day with 
an interval of 6 hours. Another bright spot on the Galactic plane and inside 
the circle is Cas A, the brightest supernova remnant in the radio sky.}
\end{figure}

Now we discuss very briefly another type of RFI from the sky due to the 
grating sidelobes of the 21CMA. Except for the easternmost pod at a distance 
of 4646 m from the center of the east-west arm, the other 80 pods that 
compose the 21CMA are not randomly deployed along the two baselines. 
Spacings between pods are integral multiples of 20 m for both 
east-west and north-south arms. 
As a result, there are only 212 independent baselines among the total 
3240 instantaneous ones. While the numerous redundant baselines 
are useful for the purpose of self-calibration and measurement of the 
EoR power spectrum (e.g. Noordam \& de Bruyn ~\cite{Noordam82}; 
Tegmark ~\cite{Tegmark10}), 
they also generate the so-called grating sidelobes 
(Bracewell \& Thompson ~\cite{Bracewell73}; 
Amy \& Large ~\cite{Amy90}). These equally spaced rings have 
their spatial responses almost comparable to that of the main lobe, 
causing serious contamination to our observation if they happen to 'see' 
the Galactic and extragalactic radio sources. For example, the two 
brightest radio sources in the sky, Cas A and Cygnus A, would both 
show up in the 21CMA field due to the grating sidelobes. All these 
troublespots for our EoR experiment should be CLEANed either from 
ungridded visibilities through a proper modeling of the sky 
or in image processing. 

Using real 21CMA real observations, 
we have thus far discussed the detection and mitigation of both 
time varying, terrestrial RFI and the time-invariant, celestial RFI. 
The next step is to move on to interferometric imaging, 
which, however, exceeds the scope of the present paper. 
The details of this can be found 
in Zheng et al. ~\cite{Zheng16}. For illustration, 
we show in Fig.13 and Fig.14 
the RFI mitigated uv map and the corresponding dirty map, respectively.
It can be seen that the spots contaminated by the Cygnus region have been 
excised in Fig.13. However, the two brightest radio sources, 
Cas A and Cygnus A, can still be found in the sky map (Fig.14), 
which are the result of the sidelobe leakage of the array. 

\begin{figure}
\centering
\includegraphics[width=10cm, angle=0]{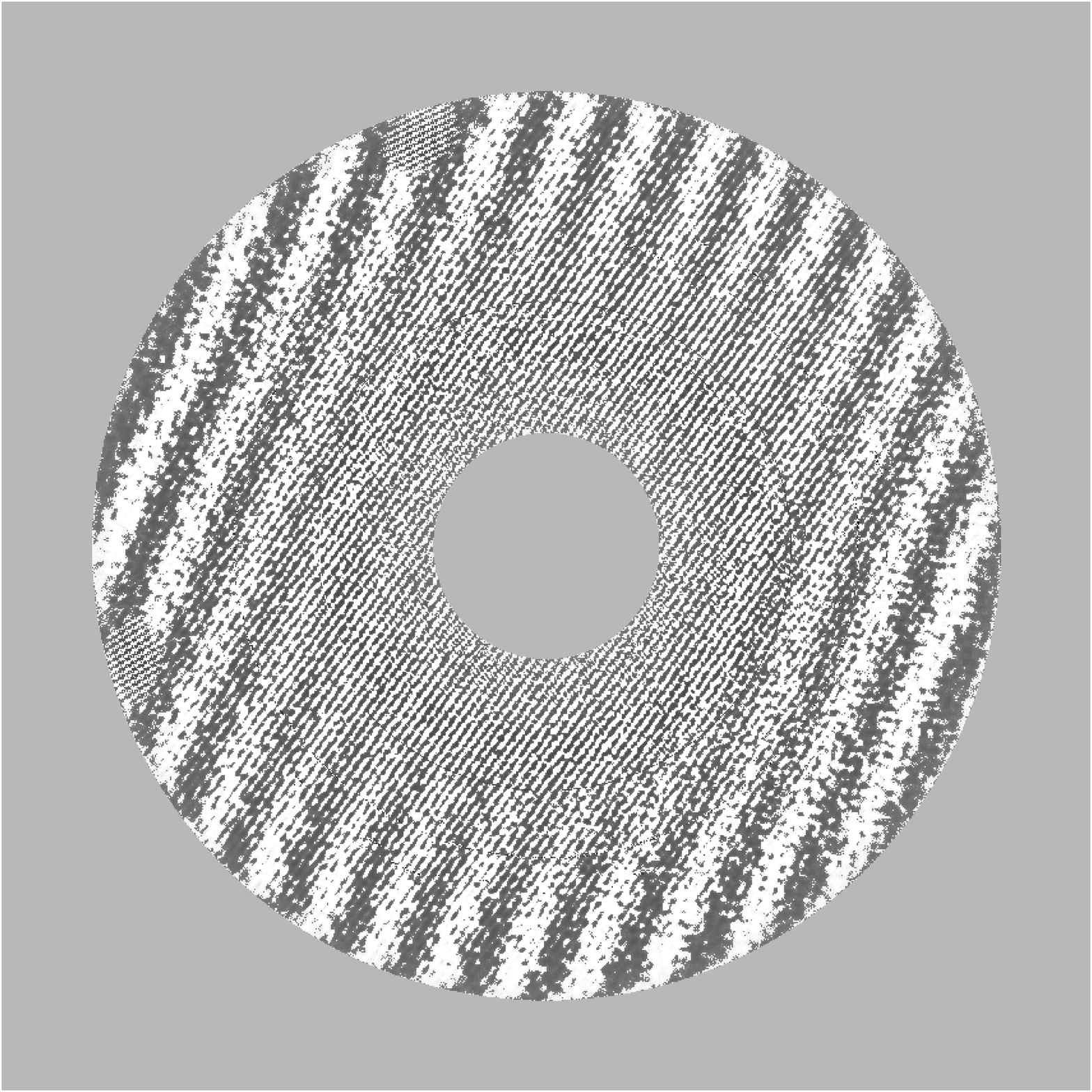}
 \caption{The 24 hour uv coverage of E01E20 after RFI is mitigated. 
Frequency runs outward from 50 MHz to 200 MHz. Two gray spots correspond 
to the sidelobe leakages of the Cygnus region in the Milky Way. 
The gray ring near 137 MHz is due to ORBCOMM satellites.}
\end{figure}

\begin{figure}
\centering
\includegraphics[width=10cm, angle=0]{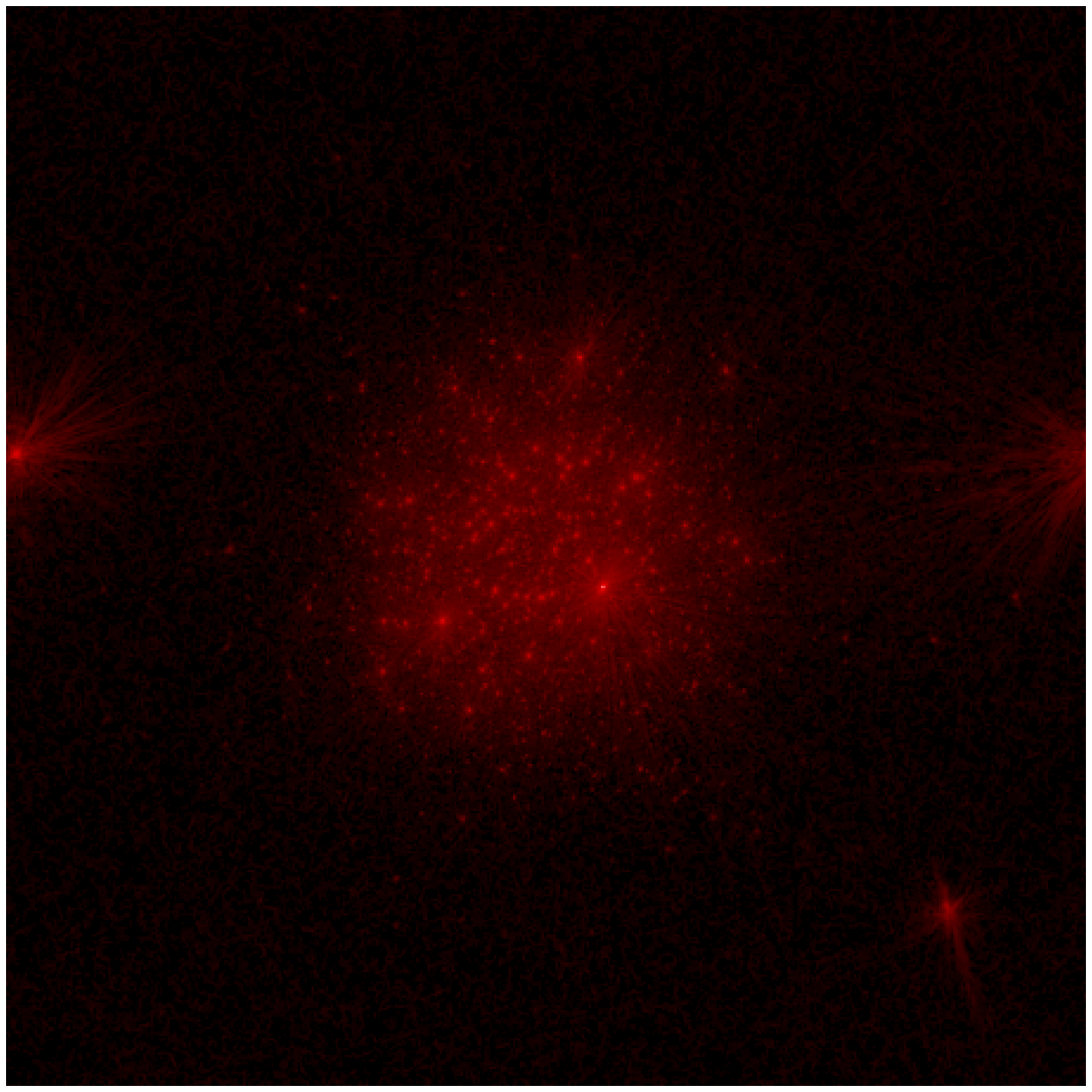}
 \caption{The dirty map of the NCP region observed with 21CMA - 
an FFT of the uv map in Fig.13. A large field of 
$60^{\circ}\times60^{\circ}$ is chosen to include the two brightest radio 
sources in the sky, Cas A (left) and Cygnus A (lower right), which 
are the result of the sidelobe leakage due to the regular 
spacings of the 21CMA antennas. No correction is made for spatial response 
function of the 21CMA antennas. }
\end{figure}

\section{Discussion and conclusions}
\label{sect:discussion}

Low frequency radio astronomy, (to be) equipped with interferometers 
like SKA and data processing using cutting-edge technologies, has entered 
a golden era of development. It will revolutionize our understanding 
of the universe including future discoveries of the dark ages, cosmic dawn 
and EoR. Low frequency astronomy will also open up a new window for 
time domain research such as pulsar timing and high precision tests 
of gravity because of its large field of view and unprecedented high 
sensitivity. Yet, these ambitious sciences are built on the promise of 
a perfect mitigation of RFI including sky noise in low frequency radio 
observations. Indeed, even if we detect and excise all the man-made, 
terrestrial RFI, the cosmological signal may still be hidden in Galactic 
and extragalactic radio backgrounds. Many sophisticated methods and 
techniques should be explored to guarantee that the sky noise can be 
suppressed to the level of $\sim1$ mK. 

Using real 21CMA observations at 50-200 MHz, we have demonstrated how RFI 
is detected and mitigated with four different approaches: 
(1) Identify and flag all the time-varying events and sparks with 
high time and frequency resolutions. It is advantageous to apply this task 
during the data acquisition phase in real time, although one can also do the 
job in data post processing to ensure the maximum efficiency of data 
acquisition. 
With the fast development and wide application of GPU and FPGA related 
technology, real-time detection of RFI with unprecedented high efficiency 
has become possible for the next generation low frequency radio facilities. 
(2) Set a proper threshold to excise all the flagged data in the 
contaminated areas and even adjacent frequency channels. 
(3) Mitigate all the non-Gaussianfeatures in terms of the criterion 
defined by mean and standard deviation. 
Using the visibility correlation coefficient instead of conventional 
visibility would improve the accuracy of RFI detection although it leads 
to a decreasing efficiency of data acquisition. 
(4) Detect and remove all bright Galactic and extragalactic radio sources 
both in the field of view and from the leakage of sidelobes. 
This last point may also be the last barrier to the detection of a 
cosmological signal from the dark ages and EoR. State-of-the-art 
algorithms developed in recent years seem to be very successful 
for removals of bright radio foregrounds.

While current low frequency radio facilities including the forthcoming SKA 
are sited in remote areas, any terrestrial radio telescopes at low 
frequencies suffer from at least two types of RFI contamination: 
(1) FM and TV broadcasting scattered by meteor trails and (2) satellite 
communications. This even does not account for the ionospheric effect on 
radio interferometric observations. 
An ideal site to do low frequency radio astronomy especially 
for the exploration of the dark ages, is the far side of the Moon. 
There are several ambitious projects that are being planned, 
either deploying radio telescopes 
on the dark side of the Moon or sending detectors into lunar orbit 
(e.g. DARE (Genova et al., ~\cite{Genova15}); 
a dedicated space-based ultra-long wavelength array 
(Rajan et al. ~\cite{Rajan15}); 
Chang'e-5; etc.). 
If one observes the radio sky from the dark 
side of the Moon, Galactic and extragalactic radiation will be the only 
sources of contamination, with time-invariant power and a 
featureless spectrum like thermal noise. Many of the present RFI detection 
and mitigation methods for highly time-varying events and sparks would 
become obsolete. That will be the next milestone for low frequency radio 
astronomy.

\begin{acknowledgements}
This work was partially supported by the National Natural Science Foundation 
of China (Grant No. 11433002). 
QZ acknowledges the support by a Marsden Fund grant in New Zealand. 
\end{acknowledgements}

\label{lastpage}

\end{document}